\pdfoutput=1
\documentclass{moriond}

\usepackage{amsmath}
\usepackage{amssymb}
\usepackage{graphicx}
\usepackage{xspace}
\usepackage{color}
\usepackage{slashed}
\usepackage{units}
\usepackage[dvipsnames]{xcolor}

\usepackage{hyperref}

\def\be{\begin{equation}}
\def\ee{\end{equation}}
\def\bea{\begin{eqnarray}}
\def\eea{\end{eqnarray}}

\def \L {\mathcal{L}} 

\def \vec#1{{\boldsymbol{#1}}}
\newcommand{\hc}{\ensuremath{\text{h.c.}}}

\newcommand{\Photo}{}

\begin{document}
\vspace*{4cm}
\title{Pati--Salam and lepton universality in B decays}

\author{Julian Heeck \footnote{Talk given at Rencontres de Moriond 2019, Electroweak Interactions and Unified Theories, La Thuile, Italy, 16--24 March 2019.}}

\address{Department of Physics and Astronomy, University of California, Irvine, California 92697-4575, USA}

\author{Daniele Teresi}

\address{Dipartimento di Fisica ``E. Fermi'', Universit\`a di Pisa,
Largo Bruno Pontecorvo 3, I-56127 Pisa, Italy\\[0.1em]INFN, Sezione di Pisa,
Largo Bruno Pontecorvo 3, I-56127 Pisa, Italy}

\maketitle\abstracts{
Recent hints for lepton-flavor non-universality in $B$-meson decays can be interpreted as hints for the existence of leptoquarks. We show that scalar leptoquarks unavoidably arise in grand unified theories, using the well-known Pati--Salam model as an example. These GUT-motivated leptoquarks can have a number of appealing features including automatic absence of proton decay, purely chiral couplings, and relations between the various leptoquark couplings. We show that $R(K^{(*)})$ can be connected to the neutrino mass matrix that arises via type-II seesaw, resulting in testable lepton flavor violation. In order to also explain $R(D^{(*)})$ one instead has to assume the existence of light right-handed neutrinos, once again with testable predictions in other $B$-meson decays and at the LHC.
}

\section{Introduction}

The strongest hints for physics beyond the Standard Model (SM) in the recent past have emerged in rare $B$-meson decays, probed by BaBar, Belle, and LHCb. Particularly clean observables are given by double ratios such as $R(K^{(*)})$ and $R(D^{(*)})$. These ratios are sensitive probes of lepton-flavor non-universality and show small but persistent deviations from their SM values. Updated results presented at this Moriond conference (see contributions by  M.~Prim,  T.~Humair and  G.~Caria) have already been included in several global fits (see Refs.~\cite{Alguero:2019ptt,Alok:2019ufo,Ciuchini:2019usw,Datta:2019zca,Aebischer:2019mlg,Kowalska:2019ley}) and have been discussed by D.~Straub (see these proceedings). We use the opportunity these proceedings offer and also update our analysis below. The difference from the published version in Ref.~\cite{Heeck:2018ntp} is however small and mainly quantitative.

New-physics explanations of $R(K^{(*)})$ ($R(D^{(*)})$) require $Z'$ ($W'$) vector bosons or leptoquarks. A popular \emph{combined} explanation uses the vector leptoquark (LQ) $U_1\sim (\vec{3}, \vec{1})_{2/3}$, which is unfortunately rather difficult to obtain in UV-complete models (see  A.~Angelescu's contribution to these proceedings). In Ref.~\cite{Heeck:2018ntp} we pointed out that \emph{scalar} LQs on the other hand can be found in abundance in grand unified theories, the simplest example being the Pati--Salam (PS) model based on the gauge group $SU(4)_{LC} \times SU(2)_L \times SU(2)_R$ (Ref.~\cite{Pati:1974yy}). The most general PS model that abandons quark--lepton mass unification actually already contains \emph{all} scalar LQs and is thus perfectly suited to explain the $B$-meson anomalies by lowering the relevant LQ masses to the TeV scale. In addition, these PS-motivated LQs do not cause proton decay, which is usually a major problem in simple SM extensions that requires the imposition of an additional proton-stabilizing symmetry based on baryon or lepton number (see Ref.~\cite{Hambye:2017qix}). Furthermore, the Pati--Salam symmetry imposes restraints on the LQ couplings and relates them to other couplings, thus providing testable predictions despite not being a one-particle combined explanation like $U_1$.
We will highlight some of these  predictions below but stress that the general idea goes beyond our examples and should be explored in other grand unified theories as well.

\section{\texorpdfstring{$R(K^{(*)})$}{R(K)} and neutrino mass}

The SM fermions as well as right-handed neutrinos reside in the PS representations
\begin{align}
\Psi_L \sim (\vec{4},\vec{2},\vec{1}) &\to \left(\vec{3},\vec{2}\right)_{\tfrac{1}{6}}\oplus\left(\vec{1},\vec{2}\right)_{-\tfrac{1}{2}} \equiv Q_L \oplus L_L\,,\\
\Psi_R \sim (\vec{4},\vec{1},\vec{2}) &\to \left(\vec{3},\vec{1}\right)_{\tfrac{2}{3}}\oplus \left(\vec{3},\vec{1}\right)_{-\tfrac{1}{3}}\oplus \left(\vec{1},\vec{1}\right)_{-1}\oplus \left(\vec{1},\vec{1}\right)_{0} \equiv u_R \oplus d_R\oplus \ell_R\oplus N_R\,,
\end{align}
which allow us to define a parity left--right exchange symmetry $\Psi_L \leftrightarrow \Psi_R$ that maximizes the predictive power of the framework and will be imposed below.
In order to break the PS group to the SM and provide masses to the right-handed neutrinos we introduce the scalar fields
\begin{align}
\Delta_L \sim (\vec{\overline{10}},\vec{3},\vec{1}) 
&\to 
\left(\vec{6},\vec{3}\right)_{-\tfrac{1}{3}}\oplus
\left(\vec{\overline{3}},\vec{3}\right)_{\tfrac{1}{3}}\oplus
\left(\vec{1},\vec{3}\right)_{1} \equiv 
\Sigma_3 \oplus S_3 \oplus \delta_3\,,\\
\Delta_R \sim (\vec{\overline{10}},\vec{1},\vec{3}) 
&\to 
\left(\vec{6},\vec{1}\right)_{-\tfrac{1}{3}}\oplus 
\left(\vec{6},\vec{1}\right)_{\tfrac{2}{3}}\oplus 
\left(\vec{6},\vec{1}\right)_{-\tfrac{4}{3}} \oplus 
\left(\vec{\overline{3}},\vec{1}\right)_{\tfrac{1}{3}}\oplus 
\left(\vec{\overline{3}},\vec{1}\right)_{-\tfrac{2}{3}}\oplus 
\left(\vec{\overline{3}},\vec{1}\right)_{\tfrac{4}{3}} \notag\\ 
&\oplus 
\left(\vec{1},\vec{1}\right)_{0}\oplus  
\left(\vec{1},\vec{1}\right)_{1}\oplus  
\left(\vec{1},\vec{1}\right)_{2} \equiv 
\Sigma_1\oplus \tilde{\Sigma}_1\oplus\overline{\Sigma}_1\oplus
S_1\oplus \tilde{S}_1\oplus\overline{S}_1\oplus
\delta_1\oplus \tilde{\delta}_1\oplus\overline{\delta}_1\,, \notag
\end{align}
which contain amongst other things all $S_j$ LQs (using the notation of Ref.~\cite{Dorsner:2016wpm}). The SM-singlet $\delta_1$ will obtain a large vacuum expectation value (VEV) that breaks PS $\to$ SM, whereas the triplet $ \delta_3$ will obtain a small SM-breaking VEV that generates type-II seesaw neutrino masses
\begin{align}\label{eq:seesaws}
M_\nu  &\simeq  - \sqrt{2} \langle \delta_3\rangle V_L^* y^L V_L^\dagger \,.
\end{align}
Here $V_L$ is the standard Cabibbo--Kobayashi--Maskawa matrix and $\langle \delta_3\rangle$ the sub-GeV triplet VEV.
Since the neutrino mixing angles and mass-squared differences are well known from neutrino-oscillation measurements, we can determine $y^L$ as a function of the unknown neutrino phases and lightest neutrino mass.
However, $y^L$ is not only the (symmetric) Yukawa coupling of $\delta_3$ to the neutrinos, but thanks to PS also the $S_3$ LQ coupling:
\begin{align}
\L = \overline{Q}_{L}^c y^L L_{L} S_3
+ \frac{1}{\sqrt{2}} \, \overline{L}_{L}^c y^L L_{L} \delta_3 + \hc
\end{align}
 The $S_3$ coupling is relevant for $R(K^{(*)})$  as it generates the desired Wilson coefficients
\begin{align}
\Delta C_9 = - \Delta C_{10} = \frac{\pi v^2}{\alpha} \,\frac{1}{V_{L,tb} V_{L,ts}^*} \, \frac{1}{M_{S_3}^2}  \, y^L_{23} y^{L*}_{22} \,.
\label{eq:C9}
\end{align}
Using the Moriond-updated results of Ref.~\cite{Aebischer:2019mlg} we require $\mathrm{Re}(\Delta C_9) \stackrel{!}{\simeq} -0.53 \pm 0.08$.
Here we are assuming all scalars except for $S_3$ to be sufficiently heavy to be irrelevant for the processes under consideration.

\begin{figure}[t]
\includegraphics[width=0.49\textwidth]{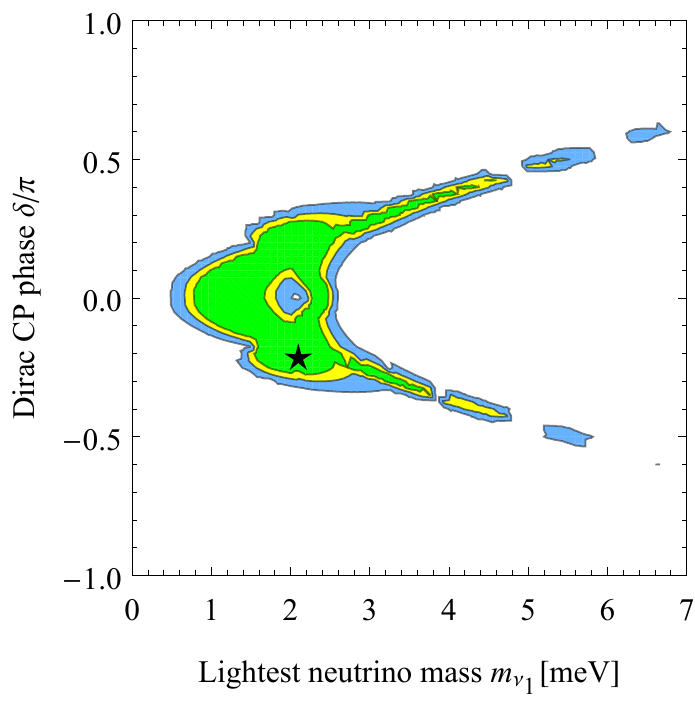}
\includegraphics[width=0.49\textwidth]{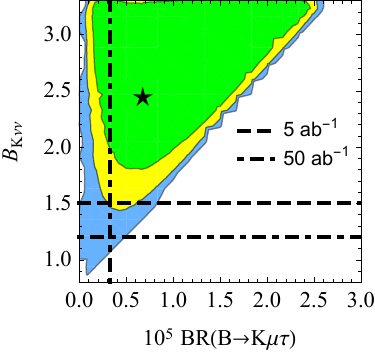}
\caption{
Left: Results of the fit for the neutral-current anomalies and type-II neutrino masses, fixing $\langle \delta_3 \rangle = 50\,$meV and normal ordering. The star denotes the best-fit point; we show the $\Delta \chi^2 < 2.3$ ($1 \sigma$), $6.2$ ($2 \sigma$), $11.8$ ($3 \sigma$) regions in green, yellow, and blue, respectively, marginalizing over all other parameters.  
Right: Results of the fit for the combined explanation of  $R(K^{(*)})$ and $R(D^{(*)})$, also showing Belle-II prospects for certain luminosities. 
}
\label{fig:type_II}
\end{figure}

PS thus relates the neutrino mass parameters of Eq.~\eqref{eq:seesaws} to the $R(K^{(*)})$ couplings of Eq.~\eqref{eq:C9}. While it is trivial to accommodate both simply by playing with $ \langle \delta_3\rangle$ and $M_{S_3}$, it is non-trivial to evade the numerous additional $S_3$-mediated processes, in particular the lepton-flavor-violating $\mu \to e $ conversion in nuclei. The existing limit of Ref.~\cite{Bertl:2006up} requires rather small $y^L_{11}$ and $y^L_{12}$ couplings compared to $y^L_{23}$ and $y^{L}_{22}$, which in turn leads to a prediction of the so-far unknown CP phases in the neutrino sector. This is in direct analogy to the often-studied texture zeros in the neutrino mass matrix, which here translate into coupling zeros of the LQ $S_3$. Fitting the unknown neutrino parameters together with $R(K^{(*)})$ allows us to make the predictions shown in Fig.~\ref{fig:type_II} (left), in particular a rather small Dirac CP phase.

We want to stress that it is highly non-trivial that the $M_\nu$--$R(K^{(*)})$ connection is successful. It only works because $M_\nu$ with normal mass ordering has a peculiar hierarchical structure: a dominant 23 block with potentially tiny  11 and 12 entries. This makes it possible for $S_3$ to explain $R(K^{(*)})$ (which resides in the 23 sector) without running into problems with muonic lepton flavor violation (which resides in the 12 sector). If we had instead assumed \emph{inverted} mass ordering we would have found an $S_3$ coupling matrix that is impossible to reconcile with $R(K^{(*)})$ and $\mu\to e$ processes. The same is true if we assume type-I seesaw dominance, as shown in Ref.~\cite{Heeck:2018ntp}.
The $M_\nu$--$R(K^{(*)})$ connection is thus a rather surprising and non-trivial feature that would be a strong hint at grand unification if confirmed. In this regard we note that the above suppression of muonic lepton flavor violation is not expected to hold far below the existing limits, so upcoming experiments such as Mu2e, COMET, and Mu3e should see signs of new physics if our explanation is correct.

\section{\texorpdfstring{$R(K^{(*)})$}{R(K)} and \texorpdfstring{$R(D^{(*)})$}{R(D)}}

While $R(K^{(*)})$ is relatively easy to accommodate in SM-extensions, $R(D^{(*)})$ requires a rather low new-physics scale around TeV and is thus subject to many more constraints. Particularly constraining is the so-far unobserved decay $B\to K\nu\nu$, which is connected to $B\to D\tau\nu$ via $SU(2)_L$ and typically requires a cancellation. As pointed out in Refs.~\cite{Becirevic:2016yqi,Asadi:2018wea,Greljo:2018ogz,Robinson:2018gza,Azatov:2018kzb}, one way out is to explain $R(D^{(*)})$ via an additional decay mode $B\to D\tau N_R$, which automatically evades the $B\to K\nu\nu$ constraint and should lead to similar differential distributions as long as $N_R$ is lighter than about $100\,$MeV.

In our PS framework the decay $B\to D\tau N_R$ can in principle be induced by the $S_1$ LQ.\footnote{An alternative solution would be to use the $R_2$ LQ that resides in the $(\vec{15},\vec{2},\vec{2})$ representation typically required for electroweak symmetry breaking. We do not consider this option here because the $R_2$ couplings are not directly related to the $S_3$ couplings, which lowers the predictivity of this approach.} However, the $S_1$ couplings are related to the $N_R$ mass matrix, in complete analogy to the $S_3$ couplings and $M_\nu$ above. This implies that a light $N_R$ is automatically weakly coupled to $S_1$, making it impossible to explain $R(D^{(*)})$. It is thus necessary to extend the PS model and decouple the $S_1$ couplings from the $N_R$ mass generation. As shown in Ref.~\cite{Heeck:2018ntp} there are numerous ways to achieve this; here we will simply assume that one of the $N_R$ is indeed light (below 100\,MeV) and has some unrelated couplings to $S_1$. Since we are staying agnostic about the neutrino mass mechanism we no longer have the $M_\nu$--$R(K^{(*)})$ connection from above, so the LQ couplings are now free parameters,
\begin{align}
\L = \overline{Q}_{L}^c y^L L_{L} S_3-  \frac{1}{\sqrt{2}} \left(\overline{d}_{R}^c y^R V_R^\dagger N_{R} + \overline{\ell}_{R}^c y^R V_R^\dagger u_{R} \right)S_1 + \hc
\end{align}
 Still, parity equates the $S_1$ and $S_3$ LQ couplings, $y^L = y^R$, for which we make the Ansatz
\begin{align}
y^L = y^R \simeq \begin{pmatrix} 
0 & 0 & 0 \\ 0 & y_{22} & y_{23} \\ 0 & y_{23} & y_{33}\end{pmatrix} 
\end{align} 
in order to evade constraints from muonic lepton flavor violation. $S_3$ will be used to explain $R(K^{(*)})$ via Eq.~\eqref{eq:C9}, i.e.~using the product of $y_{22}$ and $y_{23}$. $S_1$ will induce $B\to D\tau N_R$ via a large $y_{23}$, assuming the light $N_R$ mass eigenstate, dubbed $\hat N$, to be aligned with the second generation:
\begin{equation}\label{eq:RD}
R_{D^{(*)}} \ \equiv \ \frac{\Gamma(B \to D \tau \nu,D \tau \hat{N})/\Gamma(B \to D \tau \nu_\tau)_\text{SM}}{\Gamma(B \to D \hat{\ell} \nu,D \hat{\ell} \hat{N})/\Gamma(B \to D \hat{\ell} \nu_\ell)_\text{SM}} \, \simeq \, 1 +\left( \frac{v^2}{4 M_{S_1}^2} \frac{V_{R,cs}}{V_{L,cb}} \, y_{23}^2\right)^2  ,
\end{equation}
with $\hat{\ell}=e,\mu$, see Ref.~\cite{Heeck:2018ntp} for more details. Using the updated Belle data  on $R(D^{(*)})$ presented at this conference (and now published in Ref.~\cite{Abdesselam:2019dgh}), we estimate the preferred region to be $R_{D^{(*)}} \stackrel{!}{\simeq} 1.14 \pm 0.04$, which then determines $y_{23}/M_{S_1}$.

\begin{figure}[t]
\begin{center}
\includegraphics[width=0.6\textwidth]{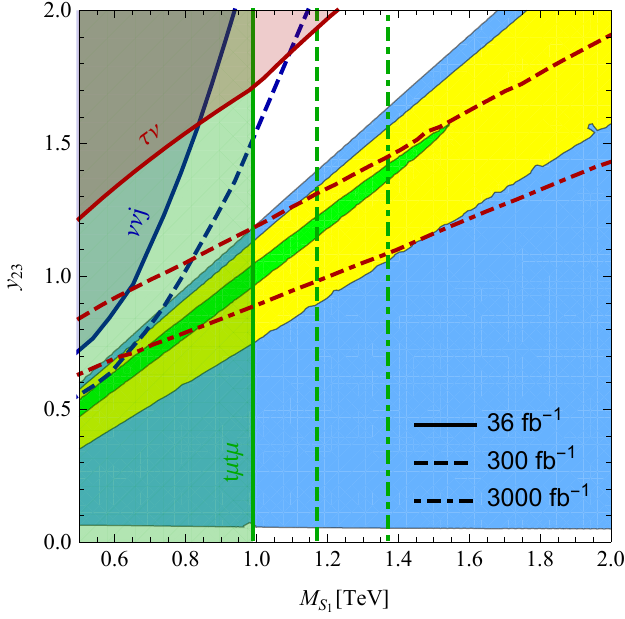}
\end{center}
\caption{
Collider phenomenology for $S_1$. The $1\sigma$, $2\sigma$, $3\sigma$ regions where both sets of anomalies are explained are denoted in green, yellow and blue, respectively. Existing LHC bounds are denoted by continuous lines and shaded regions, future prospects by dotted and dot-dashed lines for various LHC luminosities.
}
\label{fig:collider}
\end{figure}

 A combined fit of $R(K^{(*)})$ and $R(D^{(*)})$, imposing all the relevant related constraints, yields the masses and couplings of Tab.~\ref{tab:fit_combined}. Both LQs are in the TeV mass range and $y_{23}$ is of order one, much larger than all the other couplings. The main constraints and predictions then arise from processes involving the lightest LQ ($S_1$) and its 23 couplings. In particular, $B\to K\nu\nu$ and $B\to K \mu\tau$ are predicted large enough that Belle II can exclude the relevant parameter space [see Fig.~\ref{fig:type_II} (right)], while LHC can independently test this model in the channels $pp\to \tau\nu$ and $pp\to S_1 \overline{S_1}\to \bar{t}\mu t\bar{\mu}$, see Fig.~\ref{fig:collider} and Ref.~\cite{Heeck:2018ntp}. Note that we further predict the $S_1$ branching ratios $\mathrm{BR}(S_1 \to t \mu) \simeq \mathrm{BR}(S_1 \to c \tau) \simeq \mathrm{BR}(S_1 \to b \, \text{invisible})$.

\begin{table}[t]
\begin{center}
\raisebox{2.6em}{\begin{tabular}{c|c}
Parameter & Best fit \\
\hline
$M_{S_3}$ & \unit[6.1]{TeV}\\
$y_{22}$ & 0.030\\
$y_{23}$ & 1.02\\
$y_{33}$ & 0.006\\
\hline
\hline
$\chi^2$ & 2.2\\
\end{tabular}} 
\hspace{10ex}
\begin{tabular}{c|c|c}
Observable & Best fit & Pull/bound \\
\hline
$R_{D^{(*)}}$ & $1.14$ & $-0.1 \sigma$\\
$\Delta C_9 = - \Delta C_{10}$ & $-0.53$ & $+0.0\sigma$\\
$R_{D^{(*)}}^{\mu/e}$ & $1.00$ & $+0.0 \sigma$\\
$\delta g^R_{\tau \tau}$ & $0.4 \times 10^{-4}$ & $-1.3\sigma$\\
$\delta g^R_{\mu \mu}$ & $-6.9 \times 10^{-4}$ & $-0.8\sigma$\\
\hline
$B_{K\nu\nu}$ & $2.48$& $3.28$\\
$\mathrm{BR}(B \to K \mu\tau)$ & $0.7 \times 10^{-5}$ & $4.8 \times 10^{-5}$ \\
$\mathrm{BR}(\tau \to \mu \gamma)$ & $4.3 \times 10^{-11}$ & $4.4 \times 10^{-8}$\\ 
$\mathrm{BR}(\tau \to 3 \mu)$ & $7.7 \times 10^{-11}$ & $2.1 \times 10^{-8}$
\end{tabular} 
\caption{Results of the fit for the combined explanation of  $R(K^{(*)})$ and $R(D^{(*)})$, fixing $M_{S_1} = \unit[1]{TeV}$. \label{tab:fit_combined}}
\end{center}
\end{table}

\section{Conclusions}

The tantalizing and persisting $B$-meson anomalies in the observables $R(K^{(*)})$ and $R(D^{(*)})$ hint at the existence of leptoquarks, which in turn hint at grand unified theories where they arise naturally. Most grand unified theories actually contain \emph{all} scalar LQs as partners of the scalars required for the gauge-group breakdown. As a simple example we considered the well-known Pati--Salam model and showed that $R(K^{(*)})$ can be accommodated using the $S_3$ LQ and furthermore connected to the type-II seesaw neutrino mass structure, predicting measurable lepton-flavor violation in the muon sector.

Explaining both $R(K^{(*)})$ and $R(D^{(*)})$ simultaneously is more involved and requires an extension of the Pati--Salam model in order to incorporate one light right-handed neutrino with an order one coupling to the $S_1$ LQ. The underlying parity-symmetric Pati--Salam structure then identifies the $S_1$ and $S_3$ couplings and predicts large rates for $B\to K \nu\nu$ and $B\to K \mu\tau$, fully testable in Belle II. 

There is hope that the $B$-meson anomalies are just the first signs of new physics that should show up in other rare decays or even be directly produced at the LHC. Grand unified theories make for a flexible yet predictive framework that allows to connect seemingly disconnected signatures.

\section*{Acknowledgments}

JH thanks the organizers for the opportunity to present this work at Moriond and for stimulating discussions.
JH is supported, in part, by the National Science Foundation under Grant No.~PHY-1620638, and by a Feodor Lynen Research Fellowship of the Alexander von Humboldt Foundation. 
DT is supported by the ERC grant NEO-NAT.

\section*{References}

\bibliography{BIB}

\end{document}